\documentstyle[aps,multicol,epsfig]{revtex}
\begin{document}
\draft
\title{Evolution of N{\' e}el order and localized spin moment
in the doped two-dimensional Hubbard model} 
\author{Takao Morinari}
\address{Yukawa Institute for Theoretical Physics,
Kyoto University, Kyoto 606-8502, Japan}
%\address{Department of Physics, Yale University, P.O. Box 208120,
%New Haven, CT 06520-8120}
\date{\today}
\maketitle
\begin{abstract}
We investigate effects of doped holes' hopping on N{\' e}el order
in the two-dimensional Hubbard model.
Semiclassical staggered moments are computed
by solving saddle point equations derived from a path-integral
formalism.
Effects of quantum fluctuations are taken into account 
by the Schwinger boson mean field theory.
We argue that hopping of doped holes is ineffective
in suppressing N{\' e}el order
compared to rapid supprestion of N{\' e}el order 
in high-temperature superconductors.
After destruction of N{\' e}el order, the quantum disordered phase
sets in.
Taking the strong coupling limit in the quantum 
disordered phase leads to a model of spinless fermions and 
bosons but no gauge field interaction.
\end{abstract}


\begin{multicols}{2}
\narrowtext
%
% Introduction
%
\section{Introduction}
In high-temperature superconductors, one remarkable feature
is rapid destruction of N{\' e}el order by hole doping.
In fact, only 2\% doping hole concentration
is enought to suppress N{\' e}el order.
While critical disorder is 50\% for the bond percolation threshold
and 41\% for the site percolation threshold.
How do doped holes suppress N{\' e}el order in such an effective
way?

Naive expectation is that the hopping process of doped holes 
suppresses N{\' e}el oder.
In this paper, we examine the effect of disorder brought by
hopping of doped holes.
As a model, we  take the Hubbard model
because at half-filling it is reduced to the 
$S=1/2$ antiferromagnetic Heisenberg model
that describes the undoped parent compound 
of high-temperature superconductors,
and perhaps it is the simplest model to see disorder
effect by doped holes' hopping on N{\' e}el order.
We use path-integral formalism developed by Schulz.
\cite{SCHULZ,WTL}
We solve saddle point equations to compute
the magnitude of the staggered moment.
Since saddle point equations are semiclassical equations,
solutions do not contation effects of quatum fluctuations.
We take into account quantum fluctuation effects 
in terms of Schwinger boson mean field theory.\cite{AROVAS_AUERBACH}

The remainder of this paper is organized as follows:
In sec.\ref{sec_sf}, we rewrite the Hubbard model
following Schulz.\cite{SCHULZ,WTL}
In sec.\ref{sec_Neel}, we dervie saddle point equations.
Solving these equations,
we compute doping dependence of the semiclassical staggered moment.
We examine effects of quantum fluctuations on N{\' e}el order
by the Schwinger boson mean field theory.
We show that there is the quantum disordered regime 
in which quantum fluctuations suppress N{\' e}el order.
In sec.\ref{sec_qd}, we discuss the effective action
in the quantum disordered phase.
Sec. \ref{sec_dis} is devoted to summary and discussion.

%%%%%%%%%%%%%%%%%%%%%%%%%%%%%%%%%%%%%%%%%%%%%%%%%%%%%%%%%%%%%%
%
% Hubbard model
%
%%%%%%%%%%%%%%%%%%%%%%%%%%%%%%%%%%%%%%%%%%%%%%%%%%%%%%%%%%%%%%%%

\section{Path Integral formulation of the Hubbard model}
\label{sec_sf}
In order to investigate doped holes' hopping effects
on N{\' e}el order,
we first rewrite the model in a convenient form 
following Schulz. \cite{SCHULZ,WTL}
In the coherent state path-integral formulation, the partition function 
of the model is given by
${\cal Z}=\int {\cal D}\overline{c} {\cal D} c \exp (-S)$ with 
$S=S_0 + S_U$,
where
\begin{equation}
S_0 =\int_0^{\beta} d\tau \left[ 
\sum_j \overline{c}_j \left( \partial_{\tau} - \mu \right) c_j
-t\sum_{\langle i,j \rangle} \left( \overline{c}_i c_j + 
\overline{c}_j c_i
\right) 
\right],
\label{eq_S0}
\end{equation}
\begin{equation}
S_U = \int_0^{\beta} d\tau  
\sum_j U n_{j\uparrow} n_{j\downarrow}.
\end{equation}
Hereafter $\tau$ dependence of fields is implicit.
The summation $\sum_{\langle i,j \rangle}$ is taken over the
nearest neighbor sites.
Carrier fields are represented in a spinor:
$c_i = ^T \left( \begin{array}{cc}
c_{i\uparrow} & c_{i\downarrow} \end{array} \right)$ 
and 
$\overline{c}_i = \left( \begin{array}{cc}
\overline{c}_{i\uparrow} & \overline{c}_{i\downarrow}\end{array}
\right)$.
Using the identity,\cite{MORIYA} the on-site Coulomb interaction term 
can be rewritten as,
$U\sum_j n_{j\uparrow} n_{j\downarrow} = - (U/4) \sum_j
\left[ (n_{j\uparrow} + n_{j\downarrow} )^2
+ \left( \overline{c}_j 
{\mbox{\boldmath ${\bf \sigma}$}} c_j \right)^2 
+ ( n_{j\uparrow} + n_{j\downarrow} ) \right]$,
where the components of the vector 
${\mbox{\boldmath ${\bf \sigma}$}} = (\sigma_x,\sigma_y,\sigma_z)$ 
are the Pauli spin matrices.
Introducing Hubbard-Stratonovich fields for the charge and spin
fluctuations,\cite{SCHULZ,WTL}
we obtain ${\cal Z}=\int {\cal D}\overline{c}
{\cal D} c {\cal D} \tilde{\bf S} {\cal D}\phi_{c} \exp (-S_0-S_U)$, 
where,
\begin{eqnarray}
S_U &=&
\int_0^{\beta} d\tau \left[
 U \sum_j \tilde{\bf S}_j^2 
- U \sum_j 
\tilde{\bf S}_j \cdot
\overline{c}_j {\mbox{\boldmath ${\bf \sigma}$}} c_j
\right. \nonumber \\ & & \left. 
+ \frac{U}{4} \sum_j \phi_{cj}^2 
- \frac{U}{2} \sum_j \phi_{cj} \overline{c}_j c_j  \right],
\end{eqnarray}
up to constant.
Here the vector $\tilde{\bf S}_j$ represents the localized
spin moment.
The scalar $\phi_{cj}$ is associated with charge fluctuations.
For the charge degrees of freedom, 
we take the uniform value at the saddle point:
$\phi_{cj}= \langle \overline{c}_j \sigma_0 c_j \rangle 
= 1-\delta$, with $\delta$ the doped hole concentration.

Thus, the approximate action is given by
\begin{eqnarray}
S &=& S_0 + \int_0^{\beta} d\tau 
\left[
- U \sum_j 
\tilde{\bf S}_j \cdot
\overline{c}_j {\mbox{\boldmath ${\bf \sigma}$}} c_j
+ U \sum_j \tilde{\bf S}_j^2 \right].
\label{eq_Sj}
\end{eqnarray}
Note that the first term in the square brackets has the form of
Hund coupling between the localized spin moment and the carrier's
spin.

%%%%%%%%%%%%%%%%%%%%%%%%%%%%%%%%%%%%%%%%%%%%%%%%%%%%%%%%%%%%%%%%%
%   N{\' e}el order
%%%%%%%%%%%%%%%%%%%%%%%%%%%%%%%%%%%%%%%%%%%%%%%%%%%%%%%%%%%%%%%%%
\section{Evolution of N{\' e}el order by hole doping}
\label{sec_Neel}
Now we consider doped holes' hopping effects on N{\' e}el order.
Our strategy is the following:
First, we estimate semiclassical staggered magnetic moments
by solving saddle point equations.
Secondly, we examine stability of N{\' e}el order 
against quantum fluctuations.
For analysis of quantum fluctuations, we use
Schwinger boson mean field theory.\cite{AROVAS_AUERBACH}

In the N{\' e}el ordered phase, the localized spin moment
$\tilde{{\bf S}}_j$ has the following form:
\begin{equation}
\tilde{\bf S}_j = (-1)^j \tilde{S} \hat{e}_z.
\label{eq_Sjs}
\end{equation}
Substituting this into Eq.\ref{eq_Sj}, and
then performing Fourier transformations,
we obtain
\begin{eqnarray}
S &=& {\sum_{\bf k}}' \sum_{i\omega_n} 
\left(
\begin{array}{cc}
\overline{c}_k (i\omega_n) & \overline{c}_{k+Q} (i\omega_n) 
\end{array}
\right)
\nonumber \\
& & \times
\left(
\begin{array}{cc} -i \omega_n + \epsilon_k - \mu &
-U \tilde{S} \sigma_z \\
-U \tilde{S} \sigma_z & -i \omega_n - \epsilon_k - \mu
\end{array} \right)
\left( \begin{array}{c} 
c_k (i\omega_n) \\
c_{k+Q} (i\omega_n)
\end{array} \right) 
\nonumber \\
& & 
+ \beta N U \tilde{S}^2,
\end{eqnarray}
where $\epsilon_{\bf k}=-2t (\cos k_x + \cos k_y )$ and 
the summation in ${\bf k}$-space is taken over 
half of the first Brillouin-Zone. 
The energy dispersion of the carriers is given by 
$\pm E_{\bf k}$
with $E_{\bf k} = \sqrt{\epsilon_k^2+(U \tilde{S})^2}$.
After integrating out fermions, we obtain

\begin{eqnarray}
S &=& -2 {\sum_{\bf k}}'
\left\{
  \ln 
  \left[ 
 1+ {\rm e}^{ -\beta ( E_k - \mu )}
  \right]
\right. \nonumber \\ 
& & \left. + 
   \ln 
   \left[ 
 1+{\rm e}^{-\beta (-E_k - \mu ) }
   \right] 
\right\}
+ \beta N U \tilde{S}^2
\end{eqnarray}

At zero temperature,
variation of the action with respect to
the chemical potential $\mu$ yields
\begin{equation}
\frac{1}{N} \sum_{{\bf k}}
\left[ \theta (E_k +\mu) + \theta (-E_k + \mu)
\right]
= 1-\delta
\label{eq_chemi}
\end{equation}
and variation with respect to $\tilde{S}$ yields
\begin{equation}
\frac{1}{2N} \sum_{\bf k}
\frac{U}{E_k}
\left[ \theta (E_k +\mu)- \theta (-E_k + \mu)
\right] = 1,
\label{eq_tilS}
\end{equation}
where $\theta (x) =1$ for $x>0$ and zero otherwise.

\begin{figure}[htbp]
\center
\epsfxsize=2.5truein
\psfig{file=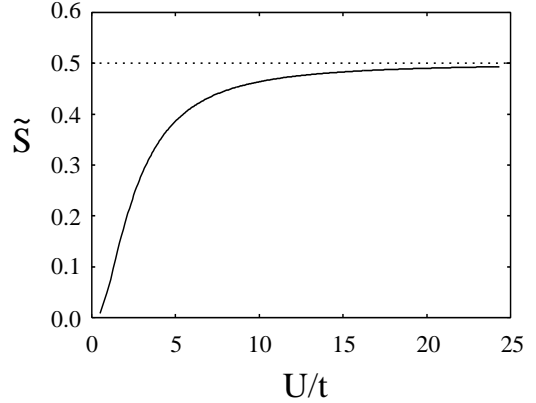,width=2.5in,angle=270}
\vspace{0.1in}
\caption{The localized spin moment $\tilde{S}$ versus $U/t$ at half-filling.}
\label{fig_half}
\end{figure}

\begin{figure}[htbp]
\center
\epsfxsize=3.2truein
\psfig{file=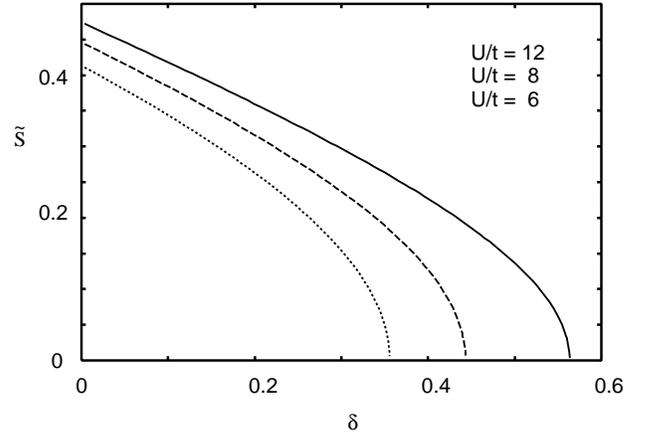,width=3.2in,angle=0}
\caption{The semiclassical staggered moment $\tilde{S}$
versus the doping concentration $\delta$ for 
$U/t=6,8,12$.}
\label{fig_utdel}
\end{figure}

\begin{figure}[htbp]
\center
\epsfxsize=3.2truein
\psfig{file=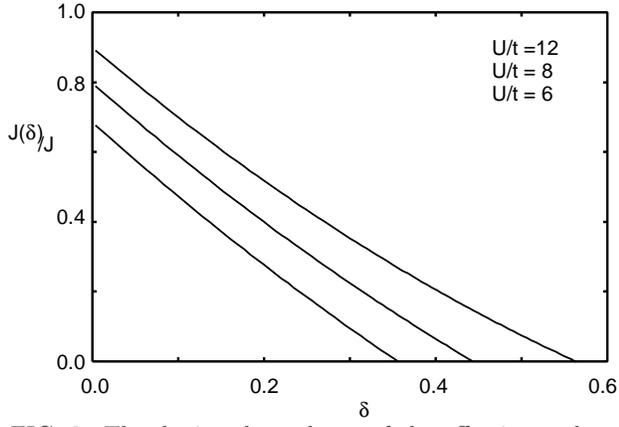,width=3.2in,angle=0}
\caption{The doping dependence of the effective 
exchange interaction $J(\delta)/J \equiv (2 \tilde{S})^2$
for $U/t=6,8,12$.
}
\label{fig_jeff}
\end{figure}

We compute $\tilde{S}$ by solving saddle point equations
(\ref{eq_chemi}) and (\ref{eq_tilS}).
Figure \ref{fig_half} shows $U/t$ dependence of
$\tilde{S}$ at half-filling.
The same result was obtained in Ref.\cite{SINGH_TESANOVIC}
based on the Hartree-Fock approximation.
Figure \ref{fig_utdel} shows doping dependence of $\tilde{S}$
at $U/t = 6,8,12$.
Figure \ref{fig_jeff} shows effective exchange interaction
$J(x)/J \equiv (2 \tilde{S})^2$.
Temperature dependence of the antiferromagnetic correlation
length is given by $\xi_{AF} = 0.26 \exp (1.38 J(x)/T)$
according to the renormalization group analysis of non-linear sigma
model and numerical simulations.\cite{MANOUSAKIS}

Note that non-zero values of $\tilde{S}$ do not necessarily
imply that there is N{\' e}el order.
Because solutions of saddle point equations neglect
quantum fluctuation effect.
We need to examine stability of N{\' e}el order
against quantum fluctuations.
For this purpose, we apply the Schwinger boson mean field theory.
\cite{AROVAS_AUERBACH}
In the Schwinger boson mean field theory,
the localized spin moment $\tilde{\bf S}_j$ is represented by
\begin{equation}
\tilde{\bf S}_j = \frac{1}{2} \overline{z}_j 
{\mbox{\boldmath ${\bf \sigma}$}}
z_j,
\label{eq_SB}
\end{equation}
with the constratint 
$\sum_{{\sigma}=\uparrow,\downarrow}z_{j\sigma}z_{j\sigma}=2\tilde{S}$.
We derive the effective action of the localized spin system by
integrating out the carrier fields.
Detail of calculations is given in Appendix \ref{ap_afh}.
The result is,
\begin{equation}
S_{\rm spin} = \int_0^{\beta} d\tau
\left[ \sum_{j\sigma} \overline{z}_{j\sigma}\partial_{\tau}
z_{j\sigma} - \frac{J}{2} \sum_{\langle i,j\rangle} 
\overline{A}_{ij} A_{ij} \right],
\end{equation}
where $J=4t^2/U$ and 
$\overline{A}_{ij}=\overline{z}_{i\uparrow} \overline{z}_{j\downarrow}
- \overline{z}_{i\downarrow} \overline{z}_{j\uparrow}$ and
$A_{ij}=z_{i\uparrow} z_{j\downarrow}
- z_{i\downarrow} z_{j\uparrow}$.
(In terms of the fields $\tilde{\bf S}_j$, $S_{\rm spin}$ is,
$S_{\rm spin} = \sum_j S_j^{\rm Berry} + \int_0^{\beta}
d\tau J \sum_{\langle i,j \rangle} 
\tilde{\bf S}_i \cdot \tilde{\bf S}_j$,
where $S_j^{\rm Berry}$ denotes the Berry phase term 
for the localized spin moment $\tilde{\bf S}_j$.
This is nothing but the 
spin $\tilde{S}$ antiferromagnetic Heisenberg model.)

\begin{figure}[htbp]
\center
\epsfxsize=2.5truein
\psfig{file=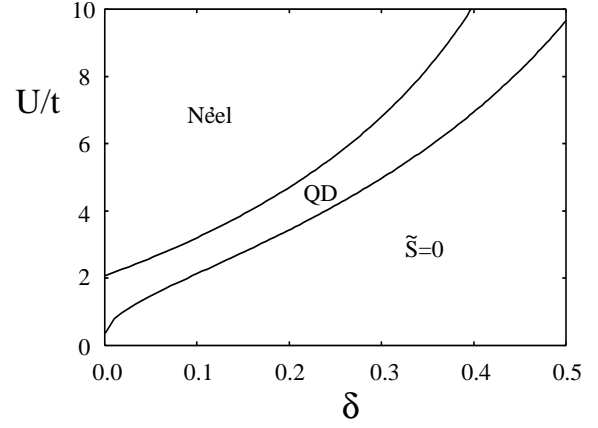,width=2.5in,angle=270}
\caption{The N{\' e}el order regime $\tilde{S}>S_c$ (N{\' e}el) 
and the quantum disordered regime 
$0< \tilde{S} < S_c$ (QD) on the $U/t$-$\delta$ plane.
In the $\tilde{S}=0$ regime, there is no antiferromagnetic correlation
that is associated with the Heisenberg antiferromagnetic type correlation.
}
\label{fig_doping}
\end{figure}

In the Schwinger boson theory, N{\' e}el order is stabilized 
if $\tilde{S}$ is larger than $S_c=0.19660$
as shown in Ref.\cite{AROVAS_AUERBACH}.
(This result is briefly summarized in Appendix \ref{ap_sc}.)
The parameter regions of $\tilde{S} > S_c$ and 
$0<\tilde{S} < S_c$ are shown in Fig. \ref{fig_doping}.
In the $\tilde{S}>S_c$ regime, N{\' e}el order is stabilized.
Whereas in the $0 < \tilde{S} < S_c$ regime, N{\' e}el order is 
suppressed by quantum fluctuations.
The point $\tilde{S} = S_c$ can be taken as 
a quantum critical point.\cite{HERTZ_MILLIS}
The $0<\tilde{S}<S_c$ regime is called the quantum disordered regime.
\cite{CHN}
Note that this quantum disordered regime
is identical to that in Ref.\cite{CHN}.
The condition of $\tilde{S} > S_c$ for the
stability of N{\' e}el order turns out to be
$g<g_c$ in Ref.\cite{CHN}.

In Fig.~\ref{fig_doping}, the $\tilde{S}=0$ regime is also shown.
Contrary to the other regimes, there is no antiferromagnetic 
Heisenberg type correlations in this regime.
Therefore, spin fluctuations in this regime are disconnected to
the original spin correlations at half-filling.

\section{Quantum disordered phase in the strong coupling limit}
\label{sec_qd}
In this section, we discuss
the effective action for the quantum disordered regime
in the strong coupling limit.
The effective action in the quantum disordered regime
is given by
\begin{eqnarray}
S &=& \int_0^{\beta} d\tau
\left[ \sum_j \overline{c}_j \left( \partial_{\tau}
- \mu \right) c_j 
- t\sum_{\langle i,j \rangle} 
\left( \overline{c}_i c_j + 
\overline{c}_j c_i \right) \right. \nonumber \\
& & \left. - \frac{U}{2}
\sum_j \left( \overline{c}_j 
{\mbox{\boldmath ${\bf \sigma}$}} c_j \right)
\left( \overline{z}_j {\mbox{\boldmath ${\bf \sigma}$}}
z_j \right)
\right. \nonumber \\
& & \left. + \sum_{j\sigma}\overline{z}_{j\sigma}
\partial_{\tau} z_{j\sigma}
+ \sum_{j\sigma} \lambda_j
\left( \overline{z}_{j\sigma} z_{j\sigma} - \tilde{S}
\right) \right. \nonumber \\
& & \left. -\frac{J}{2} \sum_{\langle i,j\rangle}
\left(\overline{z}_{j\downarrow} \overline{z}_{i\uparrow}
- \overline{z}_{j\uparrow} \overline{z}_{i\downarrow}
\right)
\left( z_{i\uparrow} z_{j\downarrow} 
- z_{i\downarrow} z_{j\uparrow} \right) \right].
\label{eq_S}
\end{eqnarray}
This action consists of the free fermion part
and the Schwinger boson part that describes
the spin $\tilde{S}$ antiferromagnetic Heisenberg model.
The interaction between them is of the form of Hund coupling.
The action (\ref{eq_S}) is so-called the spin-fermion model.

Now let us consider the strong coupling
limit, $U/t \rightarrow \infty$.
Since the Hund coupling term is dominant in this regime,
we move to a frame in which the Hund coupling term 
is diagonalized.
In such a frame, the doped hole's spin at the $j$-site
is in the direction of the localized spin moment $\tilde{\bf S}_j$.
The transformation to this frame is given by
\begin{equation}
c_j = U_j f_j,
\label{eq_tra1}
\end{equation}
where the matrix $U_j$ is given by
\begin{equation}
U_j = \left(
\begin{array}{cc} z_{j\uparrow} & -\overline{z}_{j\downarrow} \\
z_{j\downarrow} & \overline{z}_{j\uparrow} \end{array} \right).
\end{equation}
After this transformation, the action reads
\begin{eqnarray}
S
&=& \int_0^{\beta} d\tau
\left[ 
\sum_j \overline{f}_j \left(
\partial_{\tau} - \mu + \overline{U}_j \partial_{\tau} U_j \right) f_j
\right. \nonumber \\
& & \left. -t\sum_{\langle i,j \rangle} 
\left( \overline{f}_i \overline{U}_i U_j f_j
+ \overline{f}_j \overline{U}_j U_i f_i \right) \right. \nonumber \\
& & \left. 
-\frac{U}{2} \sum_j (-1)^j \overline{f}_j \sigma_z f_j \right.
\nonumber \\
& & \left.
\sum_{j\sigma}\overline{z}_{j\sigma}
\partial_{\tau} z_{j\sigma}
+ \sum_{j\sigma} \lambda_j
\left( \overline{z}_{j\sigma} z_{j\sigma} - \tilde{S}
\right) \right. \nonumber \\
& & \left. -\frac{J}{2} \sum_{\langle i,j\rangle}
\left(\overline{z}_{j\downarrow} \overline{z}_{i\uparrow}
- \overline{z}_{j\uparrow} \overline{z}_{i\downarrow}
\right)
\left( z_{i\uparrow} z_{j\downarrow} 
- z_{i\downarrow} z_{j\uparrow} \right)
\right].
\label{eq_S2}
\end{eqnarray}
Note that the hopping for the fermions from $j$-site to
$i$-site containes a matrix,
\begin{equation}
\overline{U}_i U_j = 
\left( \begin{array}{cc}
F_{ij} & -\overline{A}_{ij} \\
A_{ij} & \overline{F}_{ij} 
\end{array}
\right),
\end{equation}
with $F_{ij} = 
\overline{z}_{i\uparrow} z_{j\uparrow} 
  + \overline{z}_{i\downarrow} z_{j\downarrow}$
and $A_{ij} = 
z_{i\uparrow} z_{j\downarrow} -z_{i\downarrow} z_{j\uparrow}$.
Note that $F_{ij}$ describes ferromagnetic correlations
and $A_{ij}$ describes antiferromagnetic correlations 
in the localized spin system.
In fact, $\langle F_{ij} \rangle$ is taken for the mean field
in the forromagnetic spin system and 
$\langle A_{ij} \rangle$ is taken for the mean field
in the antiferromagnetic spin system in the 
Schwinger boson mean field theory.\cite{AROVAS_AUERBACH}.

A similar action can be derived in the slave-fermion 
mean field theory of the t-J model.
However, there is a crucial difference.
If we take the strong coupling limit of $U/t \rightarrow \infty$,
then one finds that the fermion hopping only couple to 
ferromagnetic correlations in the localized spin system.
Therefore, coupling between fermions and the gauge field 
that describes antiferromagnetic fluctuations is absent.

%%%%%%%%%%%%%%%%%%%%%%%%%%%%%%%%%%%%%%%%%%%%%%%%%%%%%%%%%%%%%%%%
%
%
\section{Summary and Discussion}
\label{sec_dis}
In this paper, we investigate effects of doped holes' hopping
on N{\' e}el order.
What we have found is that disorder effects induced by
holes' hopping on N{\' e}el order is rather small.
In fact, the critical doping concentration is
$\delta_c \simeq 0.40$ at $U/t=10$.
This value is substantially larger than that in 
high-temperature superconductors.
Therefore, hopping processes of doped holes are 
not so effective in suppressing N{\' e}el order.
For destruction of N{\' e}el order in high-temperature superconductors,
holes must behave like an excitation which suppresses
N{\' e}el order more effectively.
Such an excitation would be intimately connected with
properties of the localized spin system.

After destruction of N{\' e}el order, the quantum 
disordered phase appears.
(This quantum disordered phase is special to 
the antiferromagnetic correlations.
In fact, there is no such phase in case of 
the ferromagnetic correlations 
because there is no quantum fluctuations that suppress 
the semiclassical long-rage order as in the antiferromagnetic
case.)
If we take the strong coupling limit $U/t\rightarrow \infty$,
then the coulping between the doped holes and the antiferromagnetic
fluctuations are lost.
Here we first derive the spin-fermion model starting from
the Hubbard model.
After representing the localized spin moments by
the Schwinger bosons, we take the strong coupling limit.
There is another way of taking this strong coupling limit.
If we take the strong coupling limit first at 
the spin-fermion model, it is believed that
the model is reduced to the t-J model.
Applying the slave-boson theory, we obtain a system of 
spinless fermions and bosons with a gauge field interaction.
In the derivation of the slave-boson representation of the 
t-J model, one big assumption is that there is 
a deconfinement phase of a U(1) gauge field theory.
Whether there is a deconfinement phase or not is still 
an unsolved issue.
By contrast, there is no need to assume a deconfinement
phase when we introduce boson fields to describe 
the localized spin moments at the spin-fermion model.
After taking the strong coupling limit, we obtain
a system of spinless fermions and bosons
but there is no gauge field interaction.
Our analysis suggests that taking the strong coupling limit
is not justified or indirectly suggests
that there is no deconfinement phase.
A situation such that taking the strong coupling limit
is not allowed occurs when there is a term of
spin-orbit coupling like $H_{\rm so} 
= i\sum_{\langle i,j \rangle} c_i^{\dagger}
{\mbox{\boldmath ${\bf \lambda}$}}_{ij} \cdot
{\mbox{\boldmath ${\bf \sigma}$}} c_j + {\rm h.c.}$.
In the presence of such a term, doped holes rotate
their spin at every hopping process.\cite{MORINARI}
Therefore, we cannot take the strong coupling limit.

\acknowledgments
I thank Professor S.~Sachdev for his kind hospitality at Yale University
where part of this work was done.
This work was supported by a Grant-in-Aid from the Ministry of Education,
Culture, Sports, Science and Technology of Japan.

\appendix
\section{Derivation of the antiferromagnetic Heisenberg model}
\label{ap_afh}
In this appendix, we derive the antiferromagnetic Heisenberg model
from $S_0 + S_U$, where $S_0$ is defined by Eq.~(\ref{eq_S0})
and
$S_U$ is defined by Eq.~(\ref{eq_Sj}),
by applying second order perturbation theory with respect to $t$.
\cite{LACOUR74}.
In order to describe the localized spin moments $\tilde{\bf S}_j$,
we introduce the Schwinger bosons through Eq.\ref{eq_SB}.
with the constraint $\sum_{\sigma=\uparrow,\downarrow}
z_{j\sigma}^{\dagger} z_{j\sigma} = 2\tilde{S}_j$.
We rotate the carrier's spin in the direction of the localized spin moment 
at the same site in terms of the following unitary transformation:
\begin{equation}
c_j = U_j f_j,
\end{equation}
where
\begin{equation}
U_j = \left( 
\begin{array}{cc}
z_{j\uparrow} & -\overline{z}_{j\downarrow} \\
z_{j\downarrow} & \overline{z}_{j\uparrow} 
\end{array} \right).
\end{equation}
The total action reads
\begin{eqnarray}
S &=& \int_0^{\beta} d\tau 
\left[
\sum_j \overline{f}_j 
\left(
\partial_{\tau} - \mu + \overline{U}_j \partial_{\tau} U_j \right) f_j
\right. \nonumber \\
& & \left. -t \sum_{\langle i,j \rangle}
\left( \overline{f}_i \overline{U}_i U_j f_j
+ \overline{f}_j \overline{U}_j U_i f_i \right)
\right. \nonumber \\
& & \left. -\frac{U}{2} \sum_j \overline{f}_j \sigma_z f_j
\right].
\label{eq_a}
\end{eqnarray}

We integrate out $\overline{f}_j$ and $f_j$:
\begin{eqnarray}
S_{\rm eff} &=& -{\rm Tr} \ln \left[
\left( \partial_{\tau} - \mu + \overline{U}_j \partial_{\tau} U_j
-\frac{U}{2} \sigma_z \right) \delta_{ij} \right. \nonumber \\
& & \left. - t_{ij} \overline{U}_i U_j \right],
\label{eq_Sln}
\end{eqnarray}
where $t_{ij}=t$ for the nearest neighbor sites and 
$t_{ij}=0$ otherwise.

We expand the logarithm in Eq. (\ref{eq_Sln}) 
with respect to $t_{ij}$.
The second order term is
\begin{eqnarray}
S_{\rm eff}^{(2)} &=&
\frac12 {\rm Tr} 
\left[ \frac{1}{\partial_{\tau} - \mu
+ \overline{U}_i \partial_{\tau} U_i - \frac{U}{2} \sigma_z}
t_{ij} \overline{U}_i U_j 
\right. \nonumber \\
& & \left. \times \frac{1}{\partial_{\tau} - \mu
+ \overline{U}_j \partial_{\tau} U_j - \frac{U}{2} \sigma_z}
t_{ji} \overline{U}_j U_i  \right].
\end{eqnarray}
Applying the derivative expansion technique, we obtain
\begin{eqnarray}
S_{\rm eff}^{(2)}
&=& \frac{t^2}{\beta} 
\sum_{i\omega_n} \sum_{\langle i,j\rangle} \sum_{\sigma,\sigma'}
\frac{1}{i\omega_n + \mu + \frac{U}{2} \sigma}
\frac{1}{i\omega_n  + \mu + \frac{U}{2} \sigma'} \nonumber \\
& & \times
\int_0^{\beta} d\tau 
\langle \sigma | \overline{U}_i U_j (\tau) |\sigma' \rangle
\langle \sigma' | \overline{U}_j U_i (\tau) |\sigma \rangle
\nonumber \\
& & + ({\rm higher~derivatives})
\end{eqnarray}
After the summation over the fermion Matsubara frequencies,
we take $\beta U \rightarrow \infty$ limit.
Thus, we obtain
\begin{equation}
S_{\rm eff}^{(2)} = -\frac{J}{2} \int_0^{\beta} d\tau
\sum_{\langle i,j \rangle } \overline{A}_{ij} A_{ij},
\label{eq_sa}
\end{equation}
where $J=4t^2/U$ and 
$A_{ij} = z_{i\uparrow} z_{j\downarrow} - z_{i\downarrow} z_{j\uparrow}$
and $\overline{A}_{ij} = \overline{z}_{i\uparrow} \overline{z}_{j\downarrow}
- \overline{z}_{i\downarrow} \overline{z}_{j\uparrow}$.

On the other hand, the expansion of the logarithm
in Eq. (\ref{eq_Sln}) with respect to 
$\overline{U}_j \partial_{\tau} U_j$ gives the Berry phase term 
for the localized spin moments as follows.
The term with the first order of $\overline{U}_j \partial_{\tau} U_j$
is,
\begin{equation}
S_{\rm eff}^{\rm Berry} = -{\rm Tr} \frac{1}{\partial_{\tau}
-\mu - \frac{U}{2} \sigma_z} \overline{U}_j \partial_{\tau}
U_j.
\end{equation}
Applying the derivative expansion technique, we obtain
\begin{equation}
S_{\rm eff}^{\rm Berry} 
= \frac{1}{\beta} \sum_{i\omega_n} \sum_{\sigma}
\frac{1}{i\omega_n +\mu + \frac{U}{2} \sigma}
\sum_j \int_0^{\beta} d\tau
\langle \sigma |\overline{U}_j \partial_{\tau} U_j |\sigma
\rangle.
\end{equation}
After the summation over the fermion Matsubara frequency,
we take $\beta U \rightarrow \infty$ limit.
Thus, we obtain
\begin{equation}
S_{\rm eff}^{\rm Berry} =
\int_0^{\beta} d\tau 
\sum_{j \sigma} \overline{z}_{j\sigma} \partial_{\tau} 
z_{j\sigma}.
\end{equation}
This is nothing but the Schwinger boson representation 
of the spin's Berry phase.
In terms of the original spin moment fields $\tilde{\bf S}_j$,
the action 
$S_{\rm spin} = S_{\rm eff}^{\rm Berry} + S_{\rm eff}^{(2)}$
turns out to be the antiferromagnetic Heisenberg model.

\section{The computation of $S_c$}
\label{ap_sc}
Here we briefly summarize the result of Ref.~\cite{AROVAS_AUERBACH}
for $S_c = 0.19660$.

In terms of the Schwinger boson fields,
which is defined in Eq.~(\ref{eq_SB}), 
the antiferromagnetic Heisenberg Hamiltonian 
$H=J\sum_{\langle i,j\rangle} \tilde{\bf S}_i \cdot \tilde{\bf S}_j$,
reads
\begin{equation}
H=-\frac{J}{2} \sum_{\langle i,j\rangle}
\left( z_{i\uparrow}^{\dagger} z_{j\downarrow}^{\dagger}
-z_{i\downarrow}^{\dagger} z_{j\uparrow}^{\dagger} \right)
\left( z_{i\uparrow} z_{j\downarrow} - z_{i\downarrow} z_{j\uparrow}
\right)
+2J \tilde{S}^2 N.
\end{equation}
We introduce mean fields $A_{ij} = \langle z_{i\uparrow} z_{j\downarrow}
-z_{i\downarrow} z_{j\uparrow} \rangle$ 
and $A_{ij}^* = \langle z_{i\uparrow}^{\dagger} z_{j\downarrow}^{\dagger}
-z_{i\downarrow}^{\dagger} z_{j\uparrow}^{\dagger} \rangle$
and assume 
the uniform value $A_{ij}=A_{ij}^*=A={\rm const}$.
Then, the free energy of the system is given by
\begin{equation}
{\cal F} = \frac{2}{\beta N}
\sum_{\bf k} \ln \left[ 2\sinh \left( \frac{\beta \omega_{\bf k}}{2} \right)
\right] + J A^2 - \lambda (2\tilde{S} + 1),
\end{equation}
where $\lambda$ is a Lagrange multiplier to impose the constraint
$\sum_{\sigma} z_{j\sigma}^{\dagger} z_{j\sigma} = 2\tilde{S}$,
and $\omega_{\bf k}=\sqrt{\lambda^2-4A^2J^2 \alpha_{\bf k}^2}$
with $\alpha_{\bf k}=(\sin k_x + \sin k_y)/2$.

The variation with respect to $\lambda$ and $A$ yields
\begin{equation}
\frac{1}{N} \sum_{\bf k} \frac{J\alpha_{\bf k}^2}{\omega_{\bf k}}
\coth \frac{\beta\omega_{\bf k}}{2} = \frac12,
\end{equation}
\begin{equation}
\frac{1}{N} \sum_{\bf k} \frac{1}{\omega_{\bf k}}
\coth \frac{\beta \omega_{\bf k}}{2} =\frac{2\tilde{S}+1}{\lambda}.
\label{eq_aa2}
\end{equation}
At zero temperature, Eq.(\ref{eq_aa2}) has the following form:
\begin{equation}
\tilde{S} = \frac{1}{2}
\left( \frac{4}{\pi^2} \int_0^1 d\gamma
\frac{K(\sqrt{1-\gamma^2})}{\sqrt{1-p^2 \gamma^2}}-1 \right),
\label{eq_aa2b}
\end{equation}
with $K$ the complete elliptic function and $p=2AJ/\lambda$.
The right hand side is a monotonically increasing function 
with respect to $p$ and it takes the maximum value 
$S_c \equiv 0.19660$ at $p=1$.
Therefore, Eq.~(\ref{eq_aa2b}) has a solution of $p<1$
for $\tilde{S} < S_c$.
Solutions of $p<1$ imply that spin wave excitation 
have gap, 
and there is no Bose-Einstein condensation 
of Schwinger bosons.\cite{YOSHIOKA}.
Therefore, there is no N{\' e}el order for 
$\tilde{S}<S_c$.

%
% Figures
%1 % % % % % % % % % % % % % % % % % % % % % % % % % % % % % % % 

%2 % % % % % % % % % % % % % % % % % % % % % % % % % % % % % % % 

%3 % % % % % % % % % % % % % % % % % % % % % % % % % % % % % % % 

%4 % % % % % % % % % % % % % % % % % % % % % % % % % % % % % % % 

\end{multicols}
\end{document}